\documentclass{sf2a-conf}
\usepackage{graphicx}

\newcommand{\eq}[1]{\begin{equation} #1 \end{equation}}

\begin{document}
\TitreGlobal{SF2A 2008}
%%-----------------------------
%%      the top matter
%%-----------------------------
\title{Possible creation of net circular polarization \\ and not only depolarization of spectral lines \\  by isotropic collisions}
\address{LERMA, Observatoire de Paris -- Meudon, CNRS UMR~8112,
5~place Jules Janssen, 92195 Meudon Cedex, France}
\address{Astronomical Institute ASCR,
Fri\v{c}ova~298, 25165 Ond\v{r}ejov, Czech~Republic}
\author{\v{S}t\v{e}p\'an, J.$^{1,2}$}
\author{Sahal-Br\'echot, S.$^1$}
\runningtitle{Collisional net circular polarization}
\setcounter{page}{237} % Keep this line, even if the page will be settled afterwards..

%\author{Einstein, A.} 
%\author{Curie, W.} 
%\author{Planck, M.} 
% Repeat the authors here, this will help to make the final index

\maketitle
\begin{abstract}
We will show that isotropic collisions of electrons and protons with neutral hydrogen can lead to creation of net orientation of the atomic levels in the presence of a magnetic field. Consequently, the emitted Stokes-V profile of the spectral lines can be almost symmetric in contrast to the typical antisymmetric signature of the Zeeman effect. Moreover, the amplitude of the symmetric lobe can be significantly higher than the amplitude of the antisymmetric components. This mechanism is caused by a $\pm{M}$ symmetry breaking of the collisional transitions between different Zeeman sublevels. We will show an example of our first results for the H$\alpha$ line. This new mechanism could perhaps explain the net circular polarization of spectral lines observed in some solar limb observations and which are currently not understood. However, our results are very preliminary and more developments are needed for going further on.
\end{abstract}
%
%%-----------------------------
%%      your text
%%-----------------------------
\section{Introduction\label{sec:intro}}

%%% motivation

Recent spectropolarimetry measurements of the hydrogen lines in solar prominences made by L\'opez Ariste et al. (2005) at THEMIS have revealed an interesting property of the H$\alpha$ circular polarization profile: it has been found to be almost symmetric in most of the observations (see  Fig.1 and Fig.2 of that paper). Such a profile cannot be explained as a result of the Zeeman effect which leads to antisymmetric V-profiles.

Physically, emission of the line with net circular polarization (NCP) requires presence of atomic level orientation, i.e., imbalance of populations of the Zeeman sublevels $M$ and $-M$,
\eq{
N(\alpha JM)\neq N(\alpha J\,-M)\,.
}
Such an imbalance can be created by absorption of circularly polarized photon or by the the so-called alignment-to-orientation mechanism (Kemp et al. 1984). However, these mechanisms can hardly explain all the observations of NCP. A theoretical explanation of this enigmatic signal has been proposed by Casini \& Manso Sainz (2006). Their calculations based on the quasistatic approximation of the turbulent microscopic electric field seem to give a possible explanation of the measurements. 

However, the quasistatic approximation is not valid for interactions between the hydrogen atom and the protons or electrons of the medium in the physical conditions of  solar prominences. In fact, the proton or electron density  is so small (typically $10^{10}  \rm{c}\rm{m}^{-3}$) that the duration of the interaction is very small compared to the mean interval between two interactions, even for transitions  $nljM\to n\,l\pm 1\,j'M'$ among the fine structure levels. Consequently, it is the  impact approximation which is  valid, and not the quasisatic one.  The treatment of the interaction of hydrogen  with the protons and electrons of the medium for these transitions, which play the role for modifying the atomic polarization,  requires the theory of collisions ( Bommier et al. 1986, Sahal-Br\'echot et al. 1996). This is also the case for Stark broadening of H$\alpha$ lines at astrophysical densities (Stehl\'e et al. 1983 and further papers) .

In this work we argue for a new mechanism leading to atomic orientation. It is based on the fact that the spherical symmetry of the thermal electron-H\,{\small I} and proton-H\,{\small I} collisions is broken by the presence of the magnetic field in the prominence. We show that this can lead to a significant modification of the $\Delta n=0$ collisional transitions among the hydrogen Zeeman sublevels.

%\section{Collisional transitions $nljM\to n\,l\pm 1\,j'M'$ within the fine structure levels \label{sec:orient}}
\section{Collisional transitions \boldmath $nljM\to n\,l\pm 1\,j'M'$ within the fine structure levels \label{sec:orient}}

It is a well known fact of quantum mechanics that the transition amplitude (or probability) of the system affected by the action of an external time-dependent perturbation $V(t)$, depends on the duration of the interaction ($\tau_c$) and on the splitting of the levels ($\omega$). If the interaction is very short with respect to the level splitting ($\omega\tau_c\ll 1$) then the collisional transition amplitude is practically independent of $\omega$. This can be easily understood from the Heisenberg uncertainty relation between time and energy.  On the other hand, if the interaction time is comparable to the splitting ($\omega\tau_c\approx  1$),  the amplitude can be very sensitive to $\omega$. As a consequence it follows that a modification of the energy splitting of the levels can be used for tuning the transition probabilities in some cases. In other respects, the duration of a collision is most often very small compared to the inverse of the Larmor frequency, and thus the magnetic field can be neglected for calculating the cross-sections. In fact, up to now, they have always been calculated in zero-magnetic field in solar polarization studies. Thus the axial symmetry of the problem is conserved and only alignment  of the sublevels can be obtained. Consequently only net linear polarization can be observed.

%%% realistic scattering

Although, this is only an idealization of  realistic  problems and a more refined theory is sometimes required for explaining certain experimental situations or certain observations. Gay \& Omont (1974, 1976), Gay \& Schneider (1979), and other references therein,  have proved that the transition rates among the Zeeman sublevels of Hg ($6 ^{3}P_1$) colliding with rare gases are significantly modified by the presence of a magnetic field. They have shown that the inclusion of the Zeeman splitting in the energies of the Zeeman sublevels breaks the axial symmetry and creates orientation. Their  measurements of the $\sigma$ and  $\pi$  fluorescence intensities, emitted at right angles to the excitation direction  by a discharge lamp,  are in a good qualitative agreement with their semiclassical calculations, obtained by using a Van der Waals interaction potential : cf. for instance Fig.4 of Gay \& Omont (1976) and Fig.3 of Gay \& Schneider (1979). The cross-sections $\sigma (\alpha J 0 \to \alpha J +1)$ and $\sigma (\alpha J 0 \to \alpha J -1)$ are different, creating orientation by isotropic collisions, and the resulting circular polarization is of the same order of magnitude than the linear polarization.

%%% fine structure & Zeeman effect

The situation with hydrogen in prominences is slightly different.  First of all, we recall that the anisotropic excitation of the levels is due to the incident underlying photospheric radiation field. This only creates alignment, and thus linear polarization, due to the axial symmetry of the problem. The magnetic field is responsible for the Hanle effect, modifying the polarization degree and the direction of linear polarization. Collisions with electrons and protons of the medium depolarize the line. Only linear polarization can be obtained. This is the usual interpretation of the observations. 

Concerning collisions, it was shown in the past  by Bommier et al. (1986) that the inelastic dipolar transitions $nljM\to n\,l\pm 1\,j'M'$ among the fine structure levels due to isotropic collisions with thermal electrons and protons are the collisions which play the significant role in  hydrogen line depolarization in prominences.
The hydrogen fine structure is quasi-degenerated. It is due to a specific character of the nuclear electric field. The separation of the fine-structure levels due to spin-orbit interaction and  to the Lamb shift is very small. For $n=3$ it varies from approximately $10^{-7}$ to $10^{-5}$ eV, and it is well below the thermal energy of the 10\,000\,K particles of the prominence. The cross-sections $\sigma(nljM\to n\,l\pm 1\,j'M')$ are very high and, though of inelastic nature, can depolarize the $\rm H\alpha$ line.
A typical value of the magnetic field strength in a solar prominence is of the order of 10 G. Such a field leads to a splitting of the $nljM$ sublevels (Zeeman effect) comparable to the fine-structure splitting (see Fig.~\ref{fig1}). Considering the preceding, the question comes up in mind whether this splitting can affect the inelastic collisional rates among the Zeeman sublevels. 
Such a symmetry breaking could be responsible for conversion of the alignment and population of the levels into an orientation of the levels. A quantitative calculation shows that in the physical conditions of a prominence, the condition $\omega\tau_c\approx 1$ is satisfied for  $\Delta n=0$ and for a significant fraction of thermal electrons and especially protons. The problem to be solved is to calculate the appropriate collisional cross-sections.

%%% semiclassical limit

A first attempt to take into account the modification of the ion-hydrogen cross-sections in the case of a weak and uniform magnetic field has been done by Sakimoto (1992a,b) for applications to plasma physics. For the purpose of solar prominence diagnostics, we need to calculate more detailed cross-sections.
It is known that in the physical conditions of prominences the semiclassical limit for these collisions is usually well satisfied: the major role is played by  distant collisions of $\sim 1$\,eV particles for which
the colliding particle can be treated classically and moves along a straight path unperturbed by the hydrogen-perturber interaction.
The time-dependent electrostatic interaction potential $V(t)$ can be used and expanded to the second order. The first-order perturbation theory is  valid for calculating the collisional probability amplitudes, provided that a symmetrization of the $\it S$-matrix  is made (Seaton 1962). An integration over the relevant impact parameters and energies then gives the desired cross-sections. In our calculations we have used the first-order impact-parameter method (Seaton 1962, Sahal-Br\'echot 1969, Sahal-Br\'echot et al. 1996)  which is equivalent at the high-energy limit, to the first-order plane-wave Born approximation (Taulbjerg 1977). In fact our impact parameter method is better than the Born one for line polarization studies because we have used the improvements made by Bommier (2006) who has taken into account the momentum transfer in the calculations of the cross-sections.

%%%%%%%%%%%%

\section{Preliminary results\label{sec:concl}}
\begin{figure}[ht]
\begin{center}
\resizebox{8cm}{!}{\includegraphics  {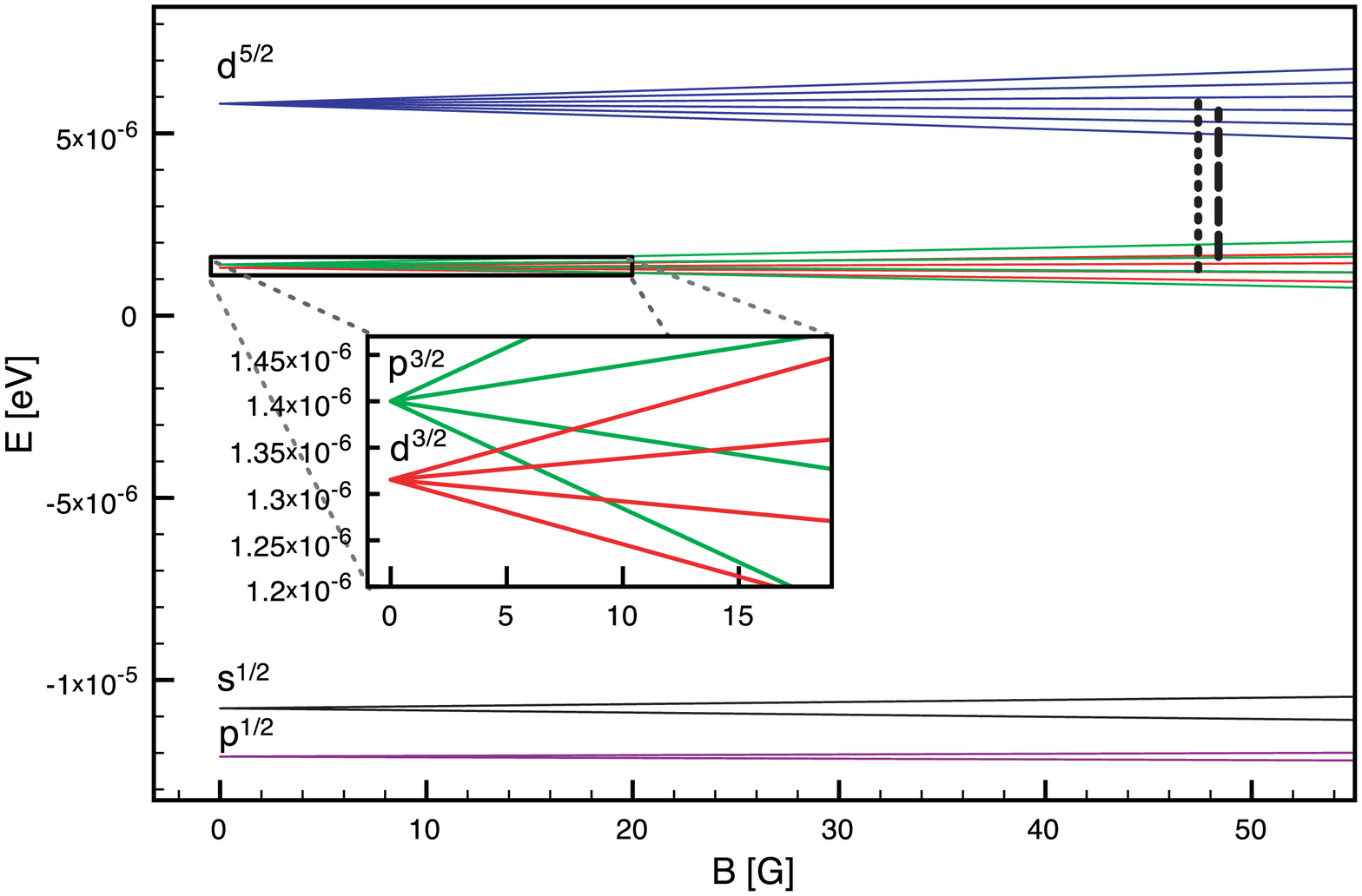}}
\resizebox{8cm}{!}{\includegraphics  {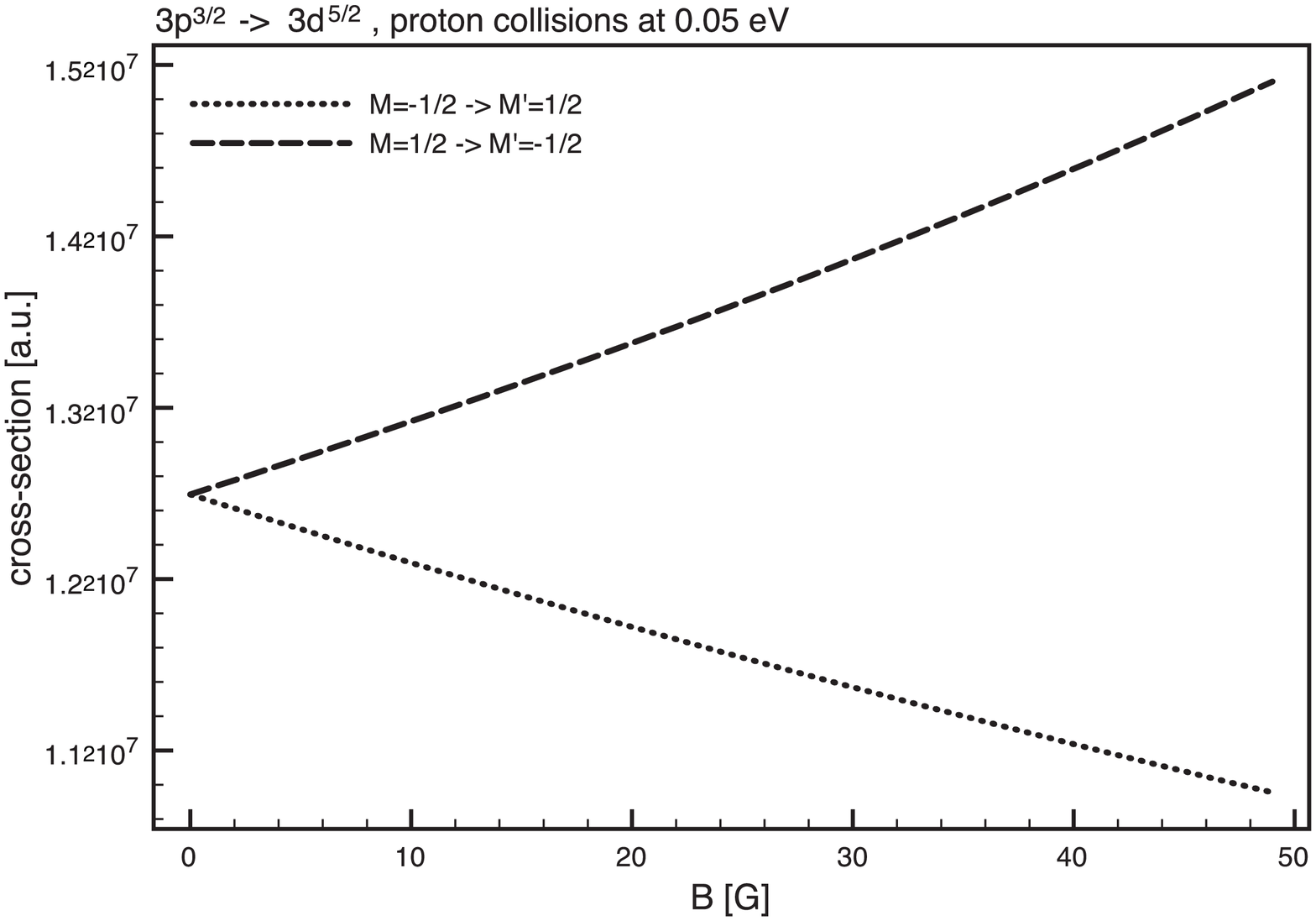}}
\caption{{\bf Left}: Zeeman splitting of the $n=3$ fine-structure levels of hydrogen.
The energy difference leads to modification of cross-sections.
{\bf Right}: Comparison of the angle-averaged proton cross-sections for $3p^{3/2}(M=-1/2)\to3d^{5/2}(M'=+1/2)$ and $3p^{3/2}(M=+1/2)\to3d^{5/2}(M'=-1/2)$ transitions at the collision energy 0.05 eV (close to the maximum cross-section value). The transitions are schematically illustrated in the left part of the figure using the same line patterns.}
\label{fig1}
\end{center}
\end{figure}

\begin{figure}[ht]
\begin{center}
\resizebox{8cm}{!}{\includegraphics  {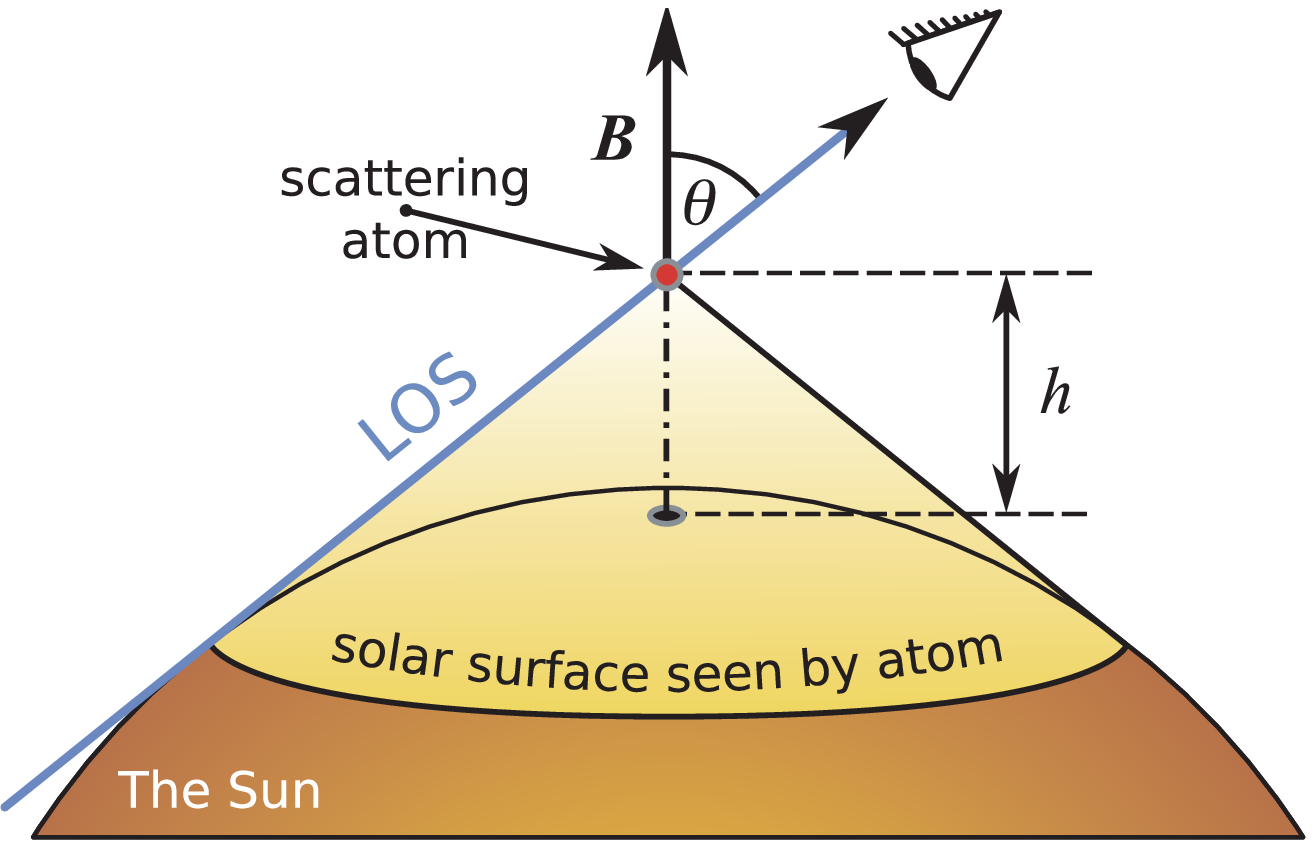}}
\resizebox{8cm}{!}{\includegraphics  {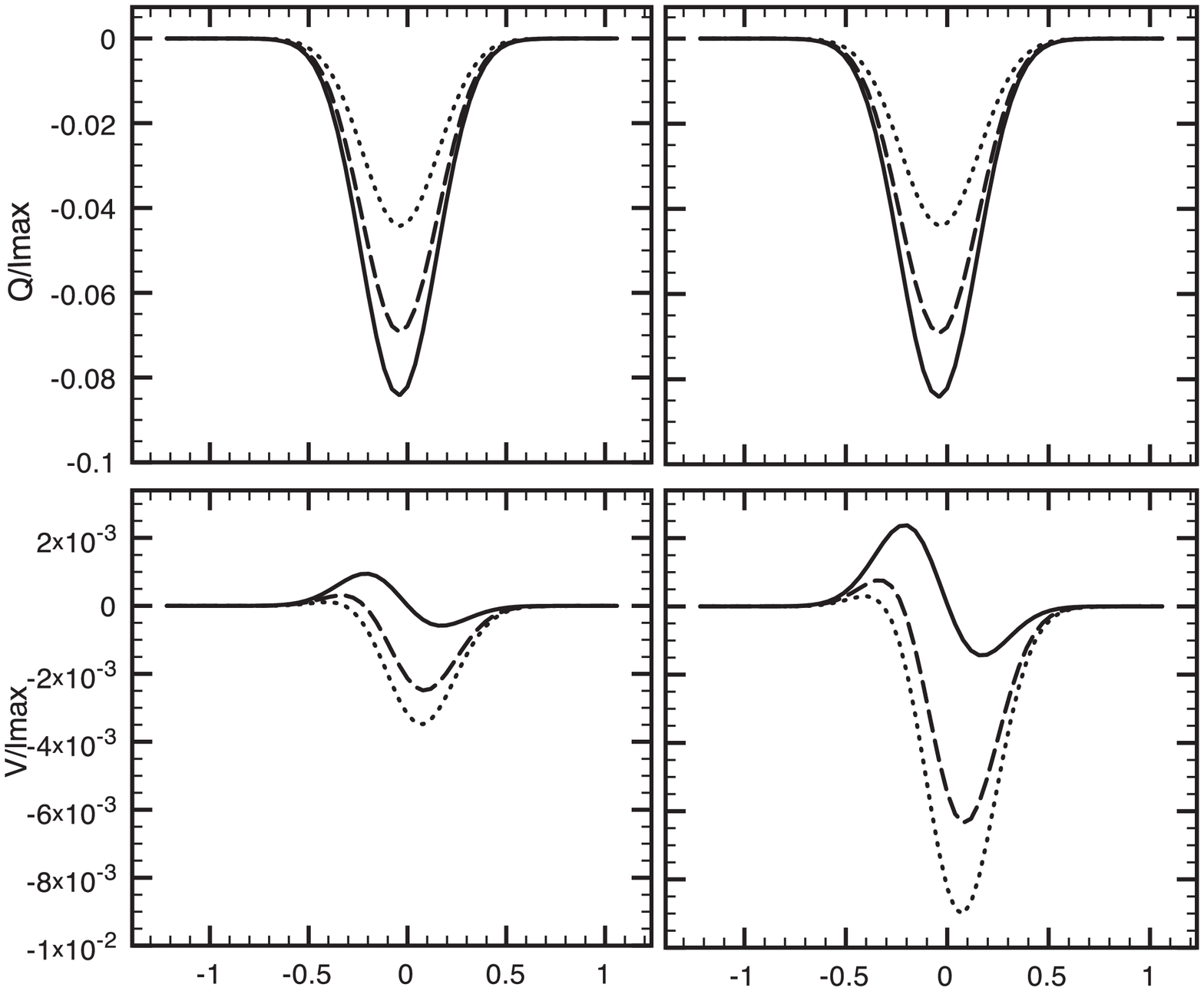}}
\caption{{\bf Left}: The geometry we consider in our example. The hydrogen atom is at the height $h$ 
above the solar surface. The magnetic field vector, $\vec B$, is vertical. The line-of-sight 
(LOS) is inclined with an angle $\theta$ from the vertical. The angle $\theta$ is 
only determined by $h/R_{\rm sun}$.
 {\bf Right}: Stokes $Q$ and $V$ profiles normalized to the maximum intensity $I_{\rm max}$ of the 
scattered H$\alpha$ radiation with a radiation anisotropy $w=0.42$ ($h=131\,000\,{\rm 
km}$, $\theta=57.3^\circ$; see Fig.~\ref{fig2}). Stokes-$U$ identically vanishes 
in our configuration.
Horizontal axis: distance from line center ($\rm \AA$).
Left (right) panels show the profiles for a magnetic field of 20~G 
(50~G). The electron and proton densities are equal: $0\,{\rm cm^{-3}}$ (solid line), $10^{10}\,{\rm cm^{-3}}$ (dashed line), and 
$5\times 10^{10}\,{\rm cm^{-3}}$ (dotted line). 
The temperature of the plasma is 8\,000~K. 
The linear polarization signal is not affected by the magnetic field in the particular considered geometry.}
\label{fig2}
\end{center}
\end{figure}

Fig.~\ref{fig1} and Fig.~\ref{fig2} show our preliminary results:  we find that this mechanism can lead to atomic orientation (i.e., net circular polarization) if the excited levels are radiatively pumped by a non-Planckian radiation.
The statistical equilibrium of the hydrogen levels in the prominence results from radiative pumping of the Zeeman sublevels by the cylindrically symmetric radiation field of the solar surface, from the Hanle effect, from the redistribution of populations and coherences of the sublevels by  thermal collisions, and by emission of radiation with thus a modified polarization signature. To keep the problem computationally simple\footnote{There are no Hanle effect and thus no quantum coherences among the Zeeman sublevels for this particular  geometry.},  we suppose that the magnetic field is oriented vertically, i.e., is parallel to the preferential direction of the incident anisotropic radiation.

% figures

The effect of isotropic collisions on creation of atomic orientation seems to be significant, even in our simple configuration. The field of few tens of Gauss and typical thermodynamical properties of prominences lead to a significant  suppression of the antisymmetric Stokes-V profile of the emitted H$\alpha$ line in favor of the symmetric lobe. An interesting property of the calculations is that the effect on the {\it linear} polarization signal is quite negligible and all the previous calculations of collisional depolarization in solar prominences remain valid.

%%%%%%%%%%%%

\section{Conclusions and future prospects\label{sec:concl}}

 We propose that this effect could be a possible explanation of the  net circular polarization observed in solar prominences.
 
The density of the perturbers  must be high enough to redistribute the radiatively unequally populated Zeeman sublevels before their radiative decay, but not too high to establish equal populations (typically $10^{10}\,{\rm cm^{-3}}$). 

In contrast to the linear polarization degree (that is always decreased by isotropic collisions), the amplitude of circular polarization can be increased in typical prominence conditions.

This mechanism does \emph{not} significantly alter the linear polarization.

More detailed calculations are needed: a realistic direction of the magnetic field must be introduced, and thus coherences have to be introduced in the calculations of the effects of collisions.

Our results are thus only preliminary and a deeper investigation of the conditions of validity of our approach also needs to be discussed.

%%-----------------------------
%%      your bibliography
%%-----------------------------
%In preparing the reference list please adhere to the following format.
% Attention should be paid to the order of the items in each reference
% and to the punctuation used. Please see Sect. 4 in the User's Guide
% that comes with the new macro package.

\end{document}